\documentclass[oldversion]{aa}

\usepackage{epsfig}
\usepackage{graphics}
\usepackage{float}
\usepackage{amsmath}
\usepackage{multirow}
\usepackage{longtable}
\usepackage{rotate}
\usepackage{array}
\usepackage{subfigure}
\def\kms{\ifmmode{\rm km\,s^{-1}}\else\hbox{$\rm km\,s^{-1}$}\fi}

\setlongtables

\begin{document}

\title{Frequentist confidence intervals for orbits}

\author{L.B.Lucy}
\offprints{L.B.Lucy}
\institute{Astrophysics Group, Blackett Laboratory, Imperial College 
London, Prince Consort Road, London SW7 2AZ, UK}
\date{Received ; Accepted }

\abstract{The problem of efficiently computing the orbital elements of a
visual binary while still deriving confidence intervals with frequentist properties
is treated. When formulated in terms of the Thiele-Innes elements, the
known distribution of probability in Thiele-Innes space allows efficient
grid-search plus Monte-Carlo-sampling schemes to be constructed for both the 
minimum-$\!\chi^{2}$ and the Bayesian approaches to parameter estimation.
Numerical experiments with 
$10^{4}$ independent realizations of an observed orbit confirm that the
$1-$ and $2\sigma$ confidence and credibility intervals have coverage
fractions close
to their frequentist values.
}

\authorrunning{Lucy}
\titlerunning{Confidence intervals}
\maketitle

\section{Introduction}

When error bars or confidence intervals are reported, the reader
expects them to have their frequentist meaning. Thus, a 95\% confidence 
interval is interpreted as implying a probability of 0.95 that the true result
is enclosed by that interval. Similarly, the interval defined by 
$\pm 1\sigma$
error bars is expected to include the true answer with probability 0.683.
However, this frequentist ideal is often not realized. 
This may
be the result of observers misjudging the precision of their
measurements or of large measurment errors occurring more frequently than
expected for a normal distribution. 

Such practical issues are absent when data analysis techniques
are investigated with simulations, since precision can be exactly specified 
and measurement errors can be assigned with random gaussian variates, so that
one might then expect a rigorous recovery of the frequentist 
ideal.
But approximations can still
compromise statistical rigour. For example, if a grid is
required, confidence intervals might be affected if the grid is too coarse.
In such cases, with increased computational resources, the limit
as the grid steps $\rightarrow 0$ can be closely approached and       
accurate results obtained.

Of more concern are approximations that compromise confidence intervals
independently
of any such limit. Two examples in the recent literature occur in {\em hybrid}
problems - i.e., non-linear problems with a subset of linear parameters.
The first example is the code EXOFAST for analysing transit and radial 
velocity data for stars with orbiting planets (Eastman et al. 2013).
These authors note that the convergence of their Markov Chain Monte Carlo
(MCMC) parameter search is much faster if the exact solution for the 
linear parameters
is introduced. However, the resulting uncertainties in the linear parameters 
are as much as 10 times smaller than when fitted non-linearly. Pending 
further research, these authors sensibly choose the inefficient option of 
treating all parameters as non-linear.

A similar but less extreme example arises when Bayesian estimation is applied
to visual binaries (Lucy 2014; L14).  
When formulated in terms of the Thiele-Innes elements, the
problem becomes linear in four of the seven elements. But when this
linearity is exploited, coverage fractions (L14, Sect.5.5)
indicate that the standard errors of the four linear elements are too small by 
factors of up to 2.1.    

These examples pose a statistical challenge in the analysis of orbits:
How can we benefit from partial linearity without losing the frequentist
properties of confidence intervals? In this paper, this challenge is
addressed in its visual binary context and for both frequentist and 
Bayesian procedures.

\section{Synthetic orbits}

The paper L14 is followed closely with regard both to notation and in the 
creation of synthetic data.

\subsection{Orbital elements} 

The orbit of the secondary relative to its primary is 
conventionally
parameterized by its  Campbell elements 
$P,e,T,a,{\rm i},\omega,\Omega$. 
Here $P$ is the period, $e$ is the eccentricity,
$T$ is a time of periastron passage, ${\rm i}$ is the inclination, 
$\omega$ is the longitude of 
periastron, and $\Omega$ is the position angle of the ascending node.
However, from the standpoint of computational economy, many investigators
- references in L14 Sect.2.1 - prefer the Thiele-Innes elements.
Thus, the Campbell parameter vector $\theta = (\phi,\vartheta)$, where
$\phi = (P,e,\tau)$ and $\vartheta = (a,{\rm i},\omega,\Omega)$, is 
replaced by the Thiele-Innes vector $(\phi,\psi)$, where
the components of the vector $\psi$ are the Thiele-Innes constants
$A,B,F,G$.   
(Note that in the $\phi$ vector,
$T$ has been replaced by $\tau = T/P$ which by definition $\in (0,1)$.)

\subsection{Model binary} 

As in L14, the adopted  model binary has the following Campbell elements:
\begin{eqnarray}
  P_{*}=100y  \;\;\; \tau_{*}=0.4 \;\;\; e_{*}=0.5  \;\;\; a_{*}=1\arcsec  
                                                   \nonumber    \\
  {\rm i}_{*}=60\degr   \;\;\;    \omega_{*} = 250\degr    \;\;\;  
                                              \Omega_{*} = 120\degr
\end{eqnarray}
An observing campaign for this binary is simulated by creating measured
Cartesian sky coordinates $(\tilde{x}_{n},\tilde{y}_{n})$ with weights
$w_{n}$ at uniformly-spaced times $t_{n}$ for $n = 1, \dots , N$ as described 
in L14 Sect.3.2. 
The parameters defining a campaign are $f_{orb}$, the fraction of the
orbit observed, $N$, the number of observations, and $\sigma$, the standard
error for unit weight. 

For given orbital elements, the predicted orbit $(x_{n},y_{n})$
is computed as described in L14 Sect.A.1, and the quality of the fit
is determined by
\begin{equation}
  \chi^{2} = \frac{1}{\sigma^{2}} \Sigma_{n} w_{n} (x_{n}-\tilde{x}_{n})^{2}
   +\frac{1}{\sigma^{2}} \Sigma_{n} w_{n} (y_{n}-\tilde{y}_{n})^{2}
\end{equation}

\section{Minimum-$\! \chi^{2}$ estimation}

The conventional (frequentist) approach to orbit-fitting is the method of
least squares - i.e., finding the elements 
$\hat{\theta} = (\hat{\phi},\hat{\psi})$ that
minimize $\chi^{2}$. When the problem is non-linear, the search for the 
minimum typically involves successive differential corrections obtained
from linearized equations, starting with an initial guess.  
However, in treating incomplete
orbits and imprecise data, it is preferable to find  
$\chi^{2}_{\min} = \chi^{2}(\hat{\theta})$ 
by means of a grid search 
(e.g., Hartkopf et al. 1989, Schaefer et al. 2006) 
and then to derive confidence intervals from constant $\chi^{2}$ 
'surfaces' in parameter space 
(e.g., Press et al. 1992, Chap. 15.6; James 2006, Chap. 9.1.2).

\subsection{Grid search}

In a brute force approach to finding $\hat{\theta}$, values of
$\chi^{2}$ would be computed throughout a 7-D grid. Confidence intervals
for the elements would then be derived from projections of the 7-D
volume ${\cal V}$ defined by the inequality 
\begin{equation}
   \chi^{2}(\theta) \: < \: \chi^{2}_{\min} +  \Delta_{\nu,\alpha}
\end{equation}
where the constant 
$\Delta_{\nu,\alpha}$ is detemined by $\nu$, the number of degrees of freedom
and $\alpha$, the desired confidence level.       
With a typical 100 steps for each dimension, the brute force method 
requires $\chi^{2}$ to be evaluated at $\sim 10^{14}$ grid points. 
However, if the linearity with respect to the Thiele-Innes elements 
can be exploited, $\chi^{2}$ is only required at $\sim 10^{6}$ grid 
points. This potential reduction by a factor of $\sim 10^{8}$ in the 
number of computed 
orbits is a powerful incentive to solve the challenge posed in Sect.1.  

On the assumption that linearity can be exploited, 
grid searches in this paper are restricted to the $\phi$-elements  
$P, e, \tau$.  
The grid is defined by taking constant steps spanning the
intervals $(\log P_{L}, \log P_{U}), (e_{L},e_{U}),(\tau_{L},\tau_{U})$. The
grid cells are labelled $(i,j,k)$ and the $\phi$-elements at the mid-points
are $\log P_{i}, e_{j}, \tau_{k}$. With these values fixed, $\chi^{2}_{ijk}$
is a function of $\psi = (A,B,F,G)$ and has its minimum value
$\hat{\chi}^{2}_{ijk}$ at the point 
$\hat{\psi}_{ijk} = (\hat{A},\hat{B},\hat{F},\hat{G})$ 
given by Eqns. (A.7) in L14.

Because $\chi^{2}_{ijk} \ge \hat{\chi}^{2}_{ijk}$,
it follows that nowhere in  
$(\phi_{ijk},\psi)$-space is  $\chi^{2}_{ijk}(\psi)$ less than the minimum value
found in the 3-D search. Accordingly, in the limit of vanishingly small
grid steps,
\begin{equation}
   \chi^{2}_{\min} =  \min_{\theta} \{\chi^{2}(\theta)\} 
                   = \min_{ijk} \: \{\hat{\chi}^{2}_{ijk}\}
\end{equation}
Thus, as has long been understood (e.g, Hartkopf et al. 1989), the 
minimum-$\!\chi^{2}$ elements $\hat{\theta}$ can be found with a grid search 
restricted to the non-linear elements.

\subsection{Approximate confidence intervals}

With $\hat{\theta}$ determined, the calculation of 
confidence intervals requires projections of ${\cal V}$, the volume
in $\theta$-space defined by Eq.(3). In the absence of a 
7-D grid, 
a possible approach is to derive approximate confidence intervals from 
projections
of the 3-D grid points satisfying Eq.(3) - i.e., from projections of the
domain ${\cal D}$ comprising grid points such that
\begin{equation}
   \hat{\chi}^{2}_{ijk}  \: < \: \chi^{2}_{\min} +  \Delta_{\nu,\alpha}
\end{equation}
This derivation of confidence intervals has exploited linearity 
since it relies on obtaining 
$\hat{\psi}_{ijk}$ and therefore  also $\hat{\chi}^{2}_{ijk}$ without iteration.
However, every point $\in {\cal D}$ also satisfies Eq.(3), 
so that ${\cal D} \in {\cal V}$. Accordingly,  
these approximate intervals will always be enclosed within the true 
intervals and so  may give a misleading impression of the elements'
precision.

\subsection{Accurate confidence intervals}

An asymptotically rigorous calculation proceeds as 
follows: first, since
$\chi^{2}_{ijk}(\psi) \ge \hat{\chi}^{2}_{ijk}$, the points  
$(\phi_{ijk},\psi)$ are all exterior to ${\cal V}$
when $\phi_{ijk} \ni {\cal D}$. Thus grid points 
$\ni {\cal D}$ are no longer of interest.

Now consider a point $\phi_{ijk} \in {\cal D}$. A point
in the $\psi$-space attached to $\phi_{ijk}$
has a $\chi^{2}$ higher than the least squares value at $\phi_{ijk}$ by the 
amount
\begin{equation}
  \delta \chi^{2}_{ijk} =  \chi^{2}_{ijk} -   \hat{\chi}^{2}_{ijk}    
\end{equation}
This point is on the 6-D surface ${\cal S}$ of the 
volume ${\cal V}$ if
\begin{equation}
  \delta \chi^{2}_{ijk} = \Delta_{\nu,\alpha}
                            -( \hat{\chi}^{2}_{ijk} - \chi^{2}_{\min}) 
\end{equation}
The contribution to ${\cal S}$ arising from the grid point $(i,j,k)$ can 
therefore
be obtained by randomly sampling the attached 
$\psi$-space subject to this constraint on $\delta \chi^{2}_{ijk}$. The
superposition of these contributions from all $\phi_{ijk}  \in {\cal D}$ then 
maps out 
${\cal S}$, and the projections of ${\cal S}$ give the desired confidence
intervals.

If ${\cal N}$ is the number of random points $\psi_{\ell}$ on ${\cal S}$ 
generated at
each $\phi_{ijk}$,
the ensemble of generated points $(\phi_{ijk},\psi_{\ell})$ becomes an exact 
representation 
of ${\cal S}$ in the limits ${\cal N} \rightarrow \infty$ and 
grid steps $\rightarrow 0$. In other words, in these limits no
finite surface element $\in {\cal S}$ would be missed by the random
sampling.

The merits of this procedure are the following: 1) The random sampling
on ${\cal S}$ does not require further orbits to be calculated; 
2) in contrast to acceptance-rejection methods common in Monte Carlo sampling,
all points are accepted;
and 3) in contrast to the brute force approach, no points are computed 
either interior or exterior to ${\cal S}$.  

\subsection{An example}

To illustrate this calculation of confidence intervals, the model binary
with elements given in Eq.(1) is observed in a campaign with parameters
$f_{orb} = 0.6, N = 15, \sigma =0\arcsec05$. The initial 3-D search 
for $\chi^{2}_{\min}$ is on a coarse $100^{3}$ grid spanning the intervals
$(1,4),(0,1),(0,1)$ in $\log P, e, \tau$, respectively. Given the resulting
initial estimate of  $\chi^{2}_{\min}$, the search boundaries are pruned
in such a way that no point with  $\chi^{2} < \chi^{2}_{\min} + 25$ is
excluded, and then a new $100^{3}$ grid is computed. The   
resulting $\chi^{2}_{\min}$ and its location are then slightly improved
by making small random displacements and acceping those that reduce $\chi^{2}$.

An investigator selects the confidence intervals of interest by specifying
$\nu$ and $\alpha$. Here we take $\alpha = 0.683$, corresponding to 
$\pm 1\sigma$ limits, and
$\nu = 1$, thus computing confidence 
intervals for each element individually - i.e., independently of the other 
elements. With these
choices, $\Delta_{\nu,\alpha} = 1$. For $\pm 2\sigma$ limits, 
$\alpha = 0.954$ and $\Delta_{\nu,\alpha} = 4$. 

With $\Delta_{\nu,\alpha} = 1$, the refined grid has 506
points satisfying Eq.(5). These define ${\cal D}$ and, as
described in Sect.3.2, approximate 
confidence intervals are obtained from projections of ${\cal D}$.
The details are as follows: At each point 
$\phi_{ijk} \in {\cal D}$, the least squares values 
$\hat{A},\hat{B},\hat{F},\hat{G}$ are available from the grid search.
From these, the Campbell elements 
$\hat{a}, \hat{\rm i}, \hat{\omega},\hat{\Omega}$ are calulated as 
described in Sect.A4 of L14. Thus the Campbell elements
$\theta_{ijk}$ are known at every point $\in {\cal D}$ and the projections of 
this ensemble give the $\pm 1\sigma$ intervals. Fiq.1 illustrates this 
procedure. 
The $(\log a, {\rm i})$-projection of the $\theta_{ijk}$ vectors 
$\in {\cal D}$ is plotted as are the resulting $\pm 1\sigma$ limits.

\begin{figure}
\vspace{8.2cm}
\includegraphics{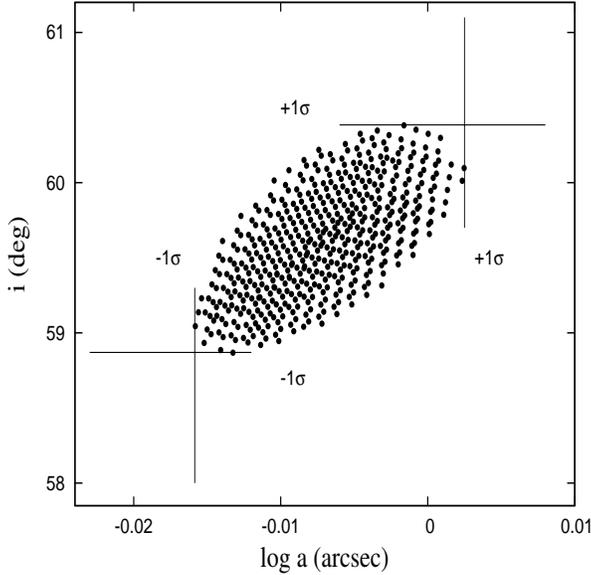}
\caption{Approximate confidence intervals. The ensemble of orbit vectors 
$\theta_{ijk} = (\phi_{ijk}, \hat{\psi}_{ijk})$ 
with $\hat{\chi}^{2}_{ijk} < \chi^{2}_{\min} + 1$ is projected on to
the $(log a, {\rm i})$-plane.
The $\pm 1 \sigma$ bounds for each coordinate 
are indicated. The 506 grid points $\phi_{ijk}$ define the domain ${\cal D}$.} 
\end{figure}

Now, for the same observed orbit $(\tilde{x}_{n}, \tilde{y}_{n})$ 
and with the same refined grid,
{\em accurate} $\pm 1\sigma$ intervals are computed according to the 
procedure of Sect.3.3.
At each of the 506 grid points $\in {\cal D}$,  ${\cal N} = 5$ random
points on ${\cal S}$ are calculated as described in Sect.(A.5). These
points $\psi_{\ell}$ are such 
that $\chi^{2}(\psi_{\ell}) = \chi^{2}_{\min} + 1$.
For each $\psi_{\ell}$, the corresponding Campbell elements are then derived as
described in Sect.A.4 of L14. The final result is 2530 vectors 
$\theta_{\ell} \in {\cal S}$. The projections of ${\cal S}$ give the
desired $\pm 1\sigma$ limits. Fig.2 illustrates this step by again projecting
the cloud of points on to the $(\log a, {\rm i})$-plane.

Because Figs.1 and 2 refer to the same orbit and are plotted to the same
scale, we see immediately that, as anticipated in Sect.3.2, the approximate
$\pm 1\sigma$ intervals are enclosed by the accurate intervals. The ratios
accurate:approximate are 1.4 for $\log a$ and 1.9 for ${\rm i}$.

A point to note from Figs. 1 and 2 is that with finite samples,
the derived confidence limits will always be underestimates. 
In the
above example, increasing ${\cal N}$ from 5 to 500 increases the
confidence intervals for $\log a$ and ${\rm i}$ by additional factors of 
1.045 and 1.055, respectively. Convergence experiments indicate that
sufficient accuracy is achieved with ${\cal N} \ga 200$.

The error bars derived with these procedures for a quantity $Q$ are in general not symmetric about its minimum-$\! \chi^{2}$ value. Accordingly, in testing
these procedures, attention is focussed not on error bars but on 
confidence intervals $(Q_{L},Q_{U})$

\begin{figure}
\vspace{8.2cm}
\includegraphics{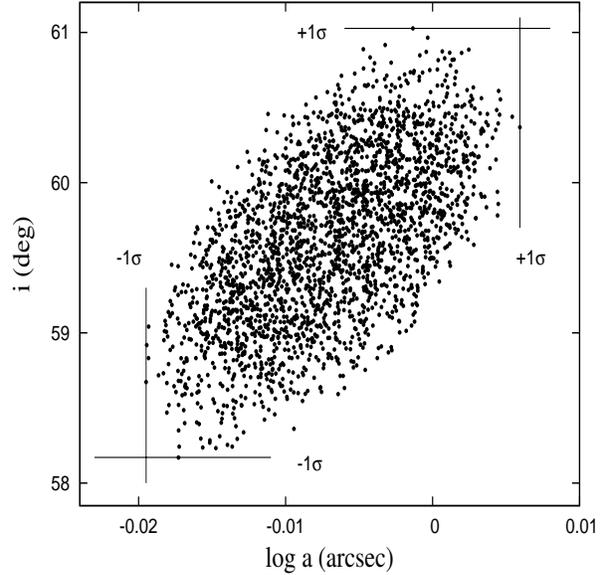}
\caption{Accurate confidence limits. The ensemble of orbit vectors $\theta_{\ell} \in {\cal S}$ - i.e., with 
$\chi^{2}_{\ell} = \chi^{2}_{\min} + 1$ - is projected on to
the $(log a, {\rm i})$-plane.
The $\pm 1 \sigma$ bounds for each coordinate 
are indicated. Each of the 506 grid points $\phi_{ijk} \in {\cal D}$ gives rise 
to ${\cal N} = 5$ points on ${\cal S}$.} 
\end{figure}

\subsection{Coverage fractions}

Confidence intervals calculated as described in Sect.3.3 are claimed to be 
asymptotically rigorous. This implies that, with a fine enough grid and a 
sufficiently large ${\cal N}$, coverage fractions will be close
to their frequentist values. To test this, two experiments similar to those 
in Sect.5.5 of L14 are now reported. In each experiment, the example of
Sect.3.4 is repeated 10,000 times with independent realizations 
$(\tilde{x}_{n}, \tilde{y}_{n})$ of the observed orbit. 
For each trial,
confidence intervals are computed for various quantities $Q$, and an
interval is counted as a success if
$Q_{L} < Q_{exact} < Q_{U}$. The quanties $Q$ are the Campbell elements
plus the mass estimator ${\cal M}$ defined in Eq.(17) of L14.

The approximate and accurate intervals
of Sect.3.2 and Sect.3.3 - now with ${\cal N} = 500$ - are investigated 
in experiments {\sc i} and {\sc ii}, respectively. In each experiment,
both $\pm 1\sigma$ and $\pm 2\sigma$ intervals are tested.

The results are reported in Table 1. From experiment {\sc i}, we see that
the approximate confidence intervals are too small
by factors of up to 1.9 for the $\pm 1\sigma$ intervals and 
up to 1.4 for the $\pm 2\sigma$ intervals. In contrast, the results for 
experiment {\sc ii} are close to ideal. Specifically, to within errors,
the $\pm 2\sigma$ coverage fractions match the frequentist value. 
On the other hand, the $\pm 1\sigma$ coverage fractions fall short by the
inconsequential factor of 1.016, indicating the need for a somewhat larger
${\cal N}$.

These results support the claim that the procedure developed in Sect.(3.3) is 
asymptotically rigorous. Moreover, the additional computational burden is
negligible: experiment {\sc ii} required a mere 11$\%$ more computer time
than experiment {\sc i}.

\begin{table}

\caption{Coverage fractions $f$ for confidence intervals from $10^{4}$ trials}

\label{table:1}

\centering

\begin{tabular}{c c c c}

\hline
\hline

 $Q$             &  expt. &   $\Delta_{\nu,\alpha} = 1  $       &  $\Delta_{\nu,\alpha} = 4  $     \\

\hline                                                                  

                                                                         \\

 ${\cal E}(f)$   &      &   $0.683 \pm 0.005$  &    $0.954 \pm 0.002$ \\ 

                                                                         \\

 $\log P $  &  {\sc i}  &   $0.627 \pm 0.005$  &  $0.948 \pm 0.002$ \\ 

            &  {\sc ii} &   $0.674 \pm 0.005$  & $0.954 \pm 0.002$  \\

 $e $       & {\sc i}   &   $0.646 \pm 0.005$  & $ 0.951 \pm 0.002$ \\ 

            & {\sc ii}  &   $0.676 \pm 0.005$  & $0.956 \pm 0.002$ \\

 $\tau $    &  {\sc i}  &   $0.639 \pm 0.005$  & $ 0.947 \pm 0.002$ \\

            & {\sc ii}  &   $0.667 \pm 0.005$  & $0.953 \pm 0.002$  \\

                                                        \\

 $\log a $  & {\sc i}   &   $0.516 \pm 0.005$  & $0.859 \pm 0.003$ \\

            & {\sc ii}  &   $0.670 \pm 0.005$  & $0.953 \pm 0.002$  \\

 ${\rm i} $  &  {\sc i}  &   $0.382 \pm 0.005$  & $ 0.691 \pm 0.005$ \\

            & {\sc ii}  &   $0.672 \pm 0.005$  & $0.952 \pm 0.002$  \\

 $\omega $  &  {\sc i}  & $0.589 \pm 0.005$    & $0.911 \pm 0.003$  \\

            & {\sc ii}  & $0.668 \pm 0.005$    & $0.954 \pm 0.002$  \\

 $\Omega $  &   {\sc i} & $0.407 \pm 0.005$    & $0.736 \pm 0.004$  \\

            &  {\sc ii} & $0.675 \pm 0.005$    & $0.955 \pm 0.002$  \\

                                                                         \\

 $\log {\cal M} $ & {\sc i} & $0.369 \pm 0.005$  & $0.684 \pm 0.005$  \\

                  &{\sc ii} & $0.672 \pm 0.005$ & $0.952 \pm 0.002$    \\

                                                               \\

\cline{1-4}

\hline
\hline

\end{tabular}

\end{table}

\section{Bayesian estimation}

In this section, the Bayesian treatment of L14 is modified to eliminate
its dependence on the profile likelihood - Eq.(3) in L14.

\subsection{Posterior density}

The posterior probability density function (pdf) at $(\phi,\psi)$  given 
data $D$ is
\begin{equation} 
 \Lambda(\phi,\psi|D)  \propto {\cal L}(\phi,\psi|D) \: \pi(\phi,\psi)
\end{equation}
where  ${\cal L}$ is the likelihood of $(\phi,\psi)$ given $D$,
and $\pi(\phi,\psi)$ is a pdf that quantifies the investigator's prior  
beliefs or knowledge about $(\phi,\psi)$. 
As in L14, we assume that 
$\pi$ is the product of seven independent priors, one for each element.
With the same choices as in L14, $\pi$ can be omitted from Eq.(8) if
the $\phi$ elements are now understood to be $(\log P, e, \tau)$.

Since coefficients independent of $(\phi, \psi)$ can be ignored
\begin{equation} 
   {\cal L}(\phi,\psi|D) \propto  \exp(-\frac{1}{2} \: \hat{\chi}^{2}) 
                     \:   \times  \:  \exp(-\frac{1}{2} \: \delta \chi^{2})   
\end{equation}
where $\hat{\chi}^{2}(\phi) = \chi^{2}(\hat{\psi}|\phi)$ is the minimum value of
$\chi^{2}$ at fixed $\phi$, and $\delta \chi^{2} (\delta \psi|\phi)$ is the 
positive 
increment in $\chi^{2}$ for the displacement $\delta \psi = \psi-\hat{\psi}$.  
The second factor in Eq.(9) can be eliminated using Eq.(A.9), so that   
\begin{equation} 
   \Lambda \: \propto \: \exp(-\frac{1}{2} \: \hat{\chi}^{2}) 
                     \:   \eta(\phi|D) \: p(\psi|\phi,D)    
\end{equation}
If we now approximate $p(\psi|\phi,D)$ by a sum of $\delta$ functions
as discussed in Sect.(A.3), then
\begin{equation} 
 \Lambda      \propto \exp(-\frac{1}{2}  \hat{\chi}^{2})    
   \:   \frac{\eta}{{\cal N}} \:  
                               \sum_{\ell} \delta(\psi-\psi_{\ell})
\end{equation}
where the
$\psi_{\ell}$ are  ${\cal N}$ independent vectors that randomly sample the
exact quadrivariate normal pdf.

The pdf $\Lambda$ is for the Thiele-Innes elements. The corresponding
pdf $\Gamma(\theta|D)$ for the Campbell elements $\theta$ is
\begin{equation} 
 \Gamma      \propto \exp(-\frac{1}{2}  \hat{\chi}^{2})    
  \:  \frac{\eta} {{\cal N}} \: 
                           \sum_{\ell} \delta(\vartheta-\vartheta_{\ell})
\end{equation}
where $\vartheta_{\ell} = \vartheta(\psi_{\ell})$.

\subsection{Credibility intervals}

In terms of a 3-D scan over $\phi$-space, the
pdf $\Gamma(\theta|D)$ is approximated by the ensemble of 7-D vectors 
\begin{equation} 
  \theta_{m} = (\phi_{ijk},\vartheta_{\ell})  
\end{equation}
with weights
\begin{equation} 
 \mu_{m}  = \frac{\eta_{ijk}}{{\cal N}} 
                           \exp(-\frac{1}{2}  \hat{\chi}^{2}_{ijk})   
\end{equation}
Here $m$ is an index that enumerates the random points
$\vartheta_{\ell}$ across all grid points $(i,j,k)$.

If $Q(\theta)$ is a quantity for which a
credibility interval is required, the data from which this can
be computed are the values $Q_{m} = Q(\theta_{m})$ with weights $\mu_{m}$.
From this data, an estimate of the pdf of $Q$ is 
\begin{equation} 
 \Theta(Q)  = \sum_{m}  \: \mu_{m} \delta(Q-Q_{m}) \:/ \: \sum_{m}  \: \mu_{m}   
\end{equation}
with corresponding cumulative distribution function (cdf)
\begin{equation} 
 F(Q)  = \sum_{Q_{m} < Q}  \: \mu_{m} \:/ \: \sum_{m}  \: \mu_{m}   
\end{equation}
The {\em equal tail} credibility interval $(Q_{L},Q_{U})$
corresponding to $\pm 1\sigma$ is then obtained from the equations
\begin{equation} 
 F(Q_{L})  = 0.1587   \;\;\;\;\;     F(Q_{U})  = 0.8413   
\end{equation}
so that the enclosed probability 0.6826.

These credibility intervals are asymptotically rigorous - i.e., are exact
in the limits ${\cal N} \rightarrow \infty$ and grid steps $\rightarrow 0$. 

\subsection{Calculation procedure}

The basic steps required to derive credibility intervals are as follows:\\
  1) At every point $(\log P_{i}, e_{j}, \tau_{k})$, the 
minimum-$\! \chi^{2}$ Thiele-Innes elements 
$\hat{A}, \hat{B},\hat{F},\hat{G}$ are obtained with Eqns.(A.7) of L14,
and the corresponding $\hat{\chi}^{2}_{ijk}$ computed.\\
  2) The variances and covariances defining the exact quadrivariate normal 
pdf $p(\psi|\phi,D)$ at $\phi_{ijk}$ are computed with Eqns. (A.9),
(A.10) of L14.\\
  3) Random points $\psi_{\ell}$  
sampling the exact pdf $p(\psi|\phi_{ijk},D)$ are computed as described 
in Sect.(A.4).\\ 
  4) The Campbell elements 
$\vartheta_{\ell}$ corresponding to
$\psi_{\ell}$ are computed as described in Sect.(A.4) of L14.\\
  5) The vectors $\theta_{m}$ are then $(\phi_{ijk},\vartheta_{\ell})$
with weights $\mu_{m}$ given by Eq.(14).\\  
  6) Lastly, credibility intervals are derived from $Q_{m}$ with the approximate
cdf given in Eq.(16).

\subsection{An example}  

To illustrate this procedure, credibility intervals are computed for the
orbit $(\tilde{x}_{n},\tilde{y}_{n})$ discussed in Sect.3.4.
A scatter diagram analogous to Figs.1 and 2 is 
not readily constructed because the points $\theta_{m}$ are not of equal
weight. Instead, the confidence intervals derived as in Fig.2 (but now with
${\cal N} = 500$) are compared with the credibility intervals derived from
Eqns.(17).

The $\Delta_{\nu,\alpha} = 1$ confidence interval for $\log a$ is
$(-0.020,0.006)$, whereas the equal-tail $68.3\%$ credibility interval is
$(-0.019,0.008)$. The corresponding intervals for $\rm i$ are
$(58\fdg1,61\fdg1)$ and $(58\fdg2,61\fdg2)$, respectively.

In these calculations, ${\cal N} = 50$ for 
$\chi^{2}_{ijk} < \chi^{2}_{min} + 21.85$ and $= 1$ otherwise. The domain
defined by this inequality corresponds to $\Delta_{\nu,\alpha}$ with
$\alpha = 0.9973$ and $\nu = 7$, thus ensuring an accurate treatment of the
wings of the posterior pdf's to beyond $\pm 2\sigma$. Convergence
experiments indicate that sufficient accuracy is achieved with 
${\cal N} \ga 20$.

In contrast to confidence intervals derived from scatter plots such as
Fig.2, the credibility intervals calculated from Eqns.(17) are not
biased. Convergence to the asymptote is therefore faster and
sufficient accuracy is achieved with a smaller ${\cal N}$.

\subsection{Coverage fractions}  

For comparison with Sect.3.5 above and with Sect.5.5 of L14, coverage
fractions for credibility intervals are computed for 10,000 independent
realizations of the observed orbit, and an interval is again counted as
a success if $Q_{L} < Q_{exact} < Q_{U}$. The results of this experiment
({\sc iii}) are given in Table 2. 

Because these are credibility not frequentist intervals, there is no
rigorous asymptotic expectation that the frequentist fractions should be
recovered. Nevertheless, these ideal fractions are closely matched and so
the credibility intervals calculated according to Sect. 4.2 can be 
described as {\em well-calibrated} (Drawid 1982).     

When Table 2 is compared to Table 1 in L14, we see that the previous
low coverage fractions for the $\psi$-elements 
$\log a, {\rm i} , \omega, \Omega$ and for the derived quantity 
$\log {\cal M}$ are now replaced by fractions close to their frequentist
values. This confirms the conjecture in L14 that the shortfall was due to
the profile likelihood.

As in Sect.3.5, statistical rigour is achieved with only a modest increase in
the computional burden. Experiment {\sc iii} required $22\%$ more 
computer time than  experiment {\sc i}.

When Bayesian estimates depend on {\em informative} priors, the pdf
$\pi(\theta)$ may have a significant gradient at $\theta_{*}$, the elements
of a particular binary. A coverage experiment restricted to  $\theta_{*}$ will
then
(correctly) deviate from the frequentist expectation. In coverage tests
for such cases, 
each independent orbit $\tilde{x}_{n},\tilde{y}_{n}$ should also be for
a random $\theta$ drawn from $\pi(\theta)$.

\begin{table}

\caption{Coverage fractions for credibility intervals from $10^{4}$ trials}

\label{table:1}

\centering

\begin{tabular}{c c c c}

\hline
\hline

 $Q$                 &  $expt.$ &     $1\sigma$       &  $ 2\sigma$      \\

\hline                                                                  
                                                                         \\

 ${\cal E}(f)$     &           &   $0.683 \pm 0.005$  &    $0.954 \pm 0.002$ \\ 

                                                                         \\

 $ \log P $  &  {\sc iii}   &   $0.664 \pm 0.005$  &  $0.950 \pm 0.002$ \\ 

 $ e $       &   {\sc iii}  &   $0.675 \pm 0.005$  & $ 0.956 \pm 0.002$ \\ 

 $ \tau $    & {\sc iii}     &   $0.680 \pm 0.005$  & $ 0.949 \pm 0.002$ \\

                                                                            \\

 $ \log a $  &  {\sc iii}    &   $0.680 \pm 0.005$  & $0.951 \pm 0.002$ \\

 ${\rm i }$       &  {\sc iii}    &   $0.685 \pm 0.005$  & $ 0.954 \pm 0.002$ \\

 $\omega $  &  {\sc iii}    & $0.682 \pm 0.005$    & $0.948 \pm 0.002$  \\

 $ \Omega $ &  {\sc iii}    & $0.677 \pm 0.005$   & $0.948 \pm 0.002$  \\

                                                                         \\

 $ \log {\cal M} $ &  {\sc iii} & $0.680 \pm 0.005$ & $0.953 \pm 0.002$  \\
                                                               \\

\cline{1-4}

\hline
\hline

\end{tabular}

\end{table}

\section{Comparison of estimates}

The relative performance of minimum-$\! \chi^{2}$ and Bayesian estimation
in experiments {\sc ii} and {\sc iii} is summarized in Table 3. The means
$<\! \delta Q \!>$ and standard deviations $s_{\delta Q}$ of the residuals 
$\delta Q = Q_{est} - Q_{exact}$
are tabulated, where $Q_{est}$ is either the minimum-$\! \chi^{2}$ value of
$Q$ or its posterior mean, and $Q_{exact}$ is given in Eq.(1).  

Table 2 shows that in this test the two
estimation methodologies yield closely similiar results. This is to be
expected for a non-informative prior $\pi(\theta)$ with negligible
gradient at $\theta_{*}$.

\begin{table}

\caption{Comparison of residuals $\delta Q$}

\label{table:2}

\centering

\begin{tabular}{c c c c c}

\hline
\hline

 $Q$            &  $<\!\delta Q \!>$ &     $ s_{\delta Q}$   & $<\!\delta Q \!>$   & $s_{\delta Q}$     \\

\hline                                                                  
                                                                           \\          
             &  $min-{\! \chi^{2}}$         &           &  Bayes       &                             \\

                                                                       \\
 $ \log P $  &  0.0021   &   0.025   &  0.0063 &  0.027 \\ 

 $ e $       &  0.0017  &   0.026   &  0.0045  & 0.027\\ 

 $ \tau $    & -0.0013     &   0.024  & -0.0044 & 0.025\\

                                                                            \\

 $ \log a $  &  0.0017    &  0.016  &  0.0041 & 0.016 \\

 ${\rm i }$  &  0\fdg079   &  1\fdg53  & 0\fdg127 & 1\fdg52\\

 $\omega $  &  -0\fdg070    & 3\fdg22    & -0\fdg37 & 3\fdg32 \\

 $ \Omega $ &  0\fdg001    & 2\fdg25   & 0\fdg133 & 2\fdg31 \\

                                                                         \\

 $ \log {\cal M} $ &  0.0010 & 0.034 & -0.0002  & 0.035 \\
                                                               \\

\cline{1-5}

\hline
\hline

\end{tabular}

\end{table}

\section{Conclusion}

This paper has addressed a technical issue in the statistical analysis of 
orbits: how to achieve statistical rigour while taking advantage of the
linearity of a subset of the orbital elements. The coverage experiments 
reported in Sects.3.5 and 4.5 demonstrate that statistical rigour
is achieved for both minimum-$\! \chi^{2}$ and Bayesian estimation.
Moreover, the reported timings show an inconsequential increase in   
the computational burden.

The key to this success is that at each grid point the distribution of 
probability in Thiele-Innes space is known. 
For minimum-$\! \chi^{2}$ estimation, this allows Monte Carlo sampling
to be targeted (Sect.A.5) 
precisely on the constant $\chi^{2}$ surface defining the desired confidence
level. For Bayesian estimation, the known pdf allows
Monte Carlo sampling in Thiele-Innes space to be concentrated (Sect.A.4)
on the high probability domain 
enclosing the least-squares points $(\hat{A},\hat{B},\hat{F},\hat{G})$.
Moreover, in each case, the random sampling does not require additional orbits
to be computed.

The approach developed here is not specific to visual binaries or to the
Thiele-Innes elements. In principle, an analogous procedure can be 
constructed for any partially linear estimation problem.

\appendix

\section{Statistics in Thiele-Innes space}

In Appendix A of L14, formulae are derived for $\hat{\psi} =
(\hat{A},\hat{B},\hat{F},\hat{G})$, the least squares  
Thiele-Innes constants at given $\phi = (\log P,e,\tau)$. The resulting value 
of $\chi^{2}$ then determines the profile likelihood ${\cal L}^{\dag}$
used in the approximation of posterior means - Eq.(7) of L14.
Now we wish to sample points displaced from $\hat{\psi}$.
Let such a displacement be $\delta \psi = (a,b,f,g)$.

\subsection{Probability density function $p(\psi|\phi,D)$}

In Sec. A.4 of L14, the pdf at $\delta \psi$ is shown to
be the product of two {\em independent} pdf's, each a
bivariate normal distribution, one for $(a,f)$, the other for  
$(b,g)$. Formulae for the variances 
$\sigma_a^{2}, \sigma_b^{2},\sigma_f^{2},\sigma_g^{2}$ and the covariances 
${\rm cov}(a,f), {\rm cov}(b,g)$ that define these pdf's are given in 
Eqs. (A.9) and (A.10) of L14.  

The pdf for $(a,f)$ is
\begin{equation}
  p(a,f) = \frac{1}{2 \pi \sigma_{a}  \sigma_{f} \sqrt{1-\rho_{af}^{2}}} 
              \: \exp( -\frac{1}{2} \xi^{2})
\end{equation}
where $\rho_{af} = {\rm cov}(a,f)/(\sigma_{a}  \sigma_{f})$ and 
\begin{equation}
  \xi^{2} = \frac{1}{1-\rho_{af}^{2}} \left [ (\frac{a}{\sigma_{a}})^{2} + 
      (\frac{f}{\sigma_{f}})^{2} - 2 \rho_{af} \: 
(\frac{a}{\sigma_{a}})(\frac{f}{\sigma_{f}})  \right ]
\end{equation}
The point $(0,0)$ corresponds to the minimum-$\! \chi^{2}_{x}$ solution
$(\hat{A},\hat{F})$, where $\chi^{2}_{x}$  is the $x$-coordinate
contribution to $\chi^{2}$
- see Eq.(2). 
The displacement $(a,f)$ therefore results in a positive increment $\delta \chi^{2}_{x}$ given by
\begin{equation}
 \sigma^{2}  \delta \chi^{2}_{x} =
 \Sigma_{n}  w_{n} \left[ (x_{n}-\tilde{x}_{n})^{2}
  -  (\hat{x}_{n}-\tilde{x}_{n})^{2} \right]
\end{equation}
Now, for displacement $(a,f)$, the predicted 
\begin{equation}
  x_{n} =  \hat{x}_{n}  +  a\: X_{n} + f \: Y_{n} 
\end{equation}
where $X,Y$ are given by Eqs. (A.3) in L14. Substitution in Eq.(A.3) then
gives
\begin{equation}
   \sigma^{2}  \delta \chi^{2}_{x} = \Sigma_{n} w_{n} \: (a^{2} X_{n}^{2} 
            + f^{2} Y_{n}^{2} + 2 \:af \:X_{n}Y_{n})
\end{equation}
where terms linear in $a$ and $f$ vanish because the minimum is at $(0,0)$.
The summations in Eq.(A.5) can be eliminated in favour of 
$\sigma_a^{2},\sigma_f^{2}$ and ${\rm cov}(a,f)$ with the formulae given in 
Eqs. (A.6), (A.9) and (A.10) of L14. After lengthy algebra, we find that
\begin{equation}
 \delta \chi^{2}_{x}   = \frac{1}{1-\rho_{af}^{2}} 
                 \left [ (\frac{a}{\sigma_{a}})^{2} + 
      (\frac{f}{\sigma_{f}})^{2} - 2 \rho_{af} \: 
(\frac{a}{\sigma_{a}})(\frac{f}{\sigma_{f}})  \right ]
\end{equation}
and so
\begin{equation}
 p(a,f) \: \propto  \: \exp ( -\frac{1}{2}  \delta \chi^{2}_{x}) 
\end{equation}  
Exactly the same analysis applies to the independent pair 
$(b,g)$, so that
\begin{equation}
 p(b,g) \: \propto  \: \exp ( -\frac{1}{2}  \delta \chi^{2}_{y}) 
\end{equation}  
Combining these formulae, we find that the pdf at 
$\psi = \hat{\psi} + \delta \psi$ is
\begin{equation}
 p(\psi| \phi,D) = \frac{1}{4 \pi^{2} \sigma^{4} \eta } \: 
\exp ( -\frac{1}{2}  \delta \chi^{2})  
\end{equation}  
where $\delta \chi^{2} =  \delta \chi^{2}_{x} +  \delta \chi^{2}_{y}$ and
$\eta(\phi|D)$ is given by
\begin{equation}
 \sigma^{4} \eta  = \sigma_{a} \sigma_{f} \sqrt(1-\rho_{af}^{2}) 
   \times \sigma_{b}\sigma_{g} \sqrt(1-\rho_{bg}^{2}) 
\end{equation}

\subsection{Modified Thiele-Innes constants}

The familiar device of 'completing the square' applied to Eq.(A.2) suggests 
new variables ${\cal A},{\cal F}$ defined by
\begin{equation}
  \frac{a}{\sigma_{a}} = {\cal A} \:,  \;\;\; 
  \frac{f}{\sigma_{f}}  = \rho_{af} \: {\cal A} + \sqrt{1-\rho_{af}^{2}} \: 
                                      {\cal F}
\end{equation}
Substitution in Eq.(A.6) then gives
\begin{equation}
 \delta \chi^{2}_{x}   =  {\cal A}^{2} + {\cal F}^{2} 
\end{equation}
The Jacobian of this transformation is
\begin{equation}
 \frac{ \partial(a,f)}{ \partial ({\cal A}, {\cal F})} =
                             \sigma_{a}  \sigma_{f} \sqrt{1-\rho_{af}^{2}}
\end{equation}
so that, by conservation of probability, the pdf 
of $({\cal A}, {\cal F})$ is
\begin{equation}
 \Psi({\cal A},{\cal F}) = \frac{1}{2 \pi} \: exp \left[ -\frac{1}{2} 
                 ({\cal A}^{2} +{\cal F}^{2}) \right] 
\end{equation}
Now, exactly the same analysis applies to the independent pair 
$(b,g)$. Thus, if we define new variables 
${\cal B}, {\cal G}$ by the equations   
\begin{equation}
  \frac{b}{\sigma_{b}} = {\cal B} \:,  \;\;\; 
  \frac{g}{\sigma_{g}}  = \rho_{bg} \: {\cal B} + \sqrt{1-\rho_{bg}^{2}} \: 
\end{equation}
then the pdf of $({\cal B},{\cal G})$ is $\Psi({\cal B},{\cal G})$, and
\begin{equation}
 \delta \chi^{2}_{y}   =  {\cal B}^{2} + {\cal G}^{2} 
\end{equation}

If $\zeta$ denotes the vector $({\cal A}, {\cal B}, {\cal F}, {\cal G})$, 
then $\Pi(\zeta)$, the 
pdf in $\zeta-$ space, is 
$\Psi({\cal A},{\cal F}) \times \Psi({\cal B},{\cal G})$ - i.e.,
\begin{equation}
 \Pi(\zeta) = \frac{1}{4 \pi^{2}} \: \exp \left[ -\frac{1}{2} 
         ({\cal A}^{2} +{\cal B}^{2} + {\cal F}^{2} +{\cal G}^{2})  \right] 
\end{equation}
Accordingly, the distribution of 
probability in $\zeta$-space is simply the product of four independent normal 
distributions, each with zero mean and unit variance. 

The increment in $\chi^{2}$ is given by Eqns.(A.12) and (A.16) as
\begin{equation}
 \delta \chi^{2}(\zeta) =  
         {\cal A}^{2} +{\cal B}^{2} + {\cal F}^{2} +{\cal G}^{2} 
\end{equation}

\subsection{Approximate pdfs}

A distribution of probability can be represented
by a sum of $\delta$ functions in such a way that the probability attached
to any {\em finite} element of space is approximated with arbitrary 
accuracy. Thus, the pdf giving the distribution 
of probability in $\zeta$-space is given approximately by
\begin{equation}
      p(\zeta) = {\cal N}^{-1} \sum_{\ell} \delta(\zeta-\zeta_{\ell})
\end{equation}  
where each $\zeta_{\ell}$ is an independent random vector sampling the
pdf $\Pi(\zeta)$ given by Eq.(A.17).
The integral of
$p(\zeta)$ over a finite element in $\zeta$-space converges to the exact value
as ${\cal N} \rightarrow \infty$.

Eqns (A.11) and (A.15) transform the point $\zeta_{\ell}$ into the displacement
$\delta \psi_{\ell} = \psi_{\ell} - \hat{\psi}$. The corresponding
approximate pdf in $\psi$-space is therefore
\begin{equation}
      p(\psi|\phi,D) = {\cal N}^{-1} \sum_{\ell} \delta(\psi-\psi_{\ell})
\end{equation}  
Note that the Jacobian of this transformation from $\zeta-$ to $\psi-$ space
is implicit in the
changes in number densities of the delta functions in the respective  
spaces.
 
Similarly, the pdf for the Campbell elements $\vartheta$ corresponding
to the Thiele-Innes elements $\psi$ is
\begin{equation}
      p(\vartheta|\phi,D) = {\cal N}^{-1} \sum_{\ell} 
                      \delta(\vartheta-\vartheta_{\ell})
\end{equation}  
where $\vartheta_{\ell} = \vartheta(\psi_{\ell})$ is derived  as described in
Sect.(A.4) of L14.

\subsection{Random sampling in $\psi$-space}

According to Eq.(A.17), a random point in 
$\zeta$-space is $(z_{1},z_{2},z_{3},z_{4})$,
where the $z_{i}$ are independent random gaussian variates drawn from ${\cal N}(0,1)$.
This point corresponds to the displacement $(a,b,f,g)$ given by Eqns.(A.11) and
(A.15) and therefore to the point 
$(\hat{A} +a, \hat{B} +b,\hat{F} +f,\hat{G} +g)$ in $\psi$-space.
Thus, a point randomly selected from the exact pdf 
$p(\psi|\phi,D)$ can be derived from 
four independent gaussian variates, and the resulting increment in
$\chi^{2}$ is given by Eq.(A.18) as
\begin{equation}
 \delta \chi^{2}(\psi)   = z_{1}^{2} + z_{2}^{2} + z_{3}^{2} + z_{4}^{2}
\end{equation}  

\subsection{Random sampling at fixed $\delta \chi^{2}$}

A random point in $\psi$-space subject to
a constraint on $\delta \chi^{2}$ can be found by first selecting a random
point on the 4-D sphere in $\zeta$-space defined by Eq.(A.18). This is achieved as follows: 
If $z_{i}$ again denotes a gaussian variate from ${\cal N}(0,1)$, then 
a random point
on this hypersphere is  
$(z_{1},z_{2},z_{3},z_{4})/Z$, where 
\begin{equation}
 Z^{2}  = (z_{1}^{2} + z_{2}^{2} + z_{3}^{2} + z_{4}^{2})/\delta \chi^{2}
\end{equation}  
(Muller 1979). The corresponding point $(A,B,F,G)$  
in $\psi$-space is then derived from Eqns.(A.11)
and (A.15). \\

The random sampling procedures of Sects.(A.4) and (A.5) {\em predict}
$\chi^{2}$ without the need to compute an orbit. This is achieved by
exploiting the linearity of the $\psi$-elements and is the basis of
the computational efficiency of the techniques of Sects.(3.3) and (4.1).
However, during code development, this prediction 
should be tested by actually computing the orbit and independently
evaluating $\chi^{2}$ from Eq.(2).

\acknowledgement

The issue of error underestimation in hybrid problems 
was raised by the referee of the previous paper (L14) and was the direct
stimulus of this investigation. This same referee
provided useful comments on this paper.


\begin{thebibliography}{}


\bibitem[]{}

 Dawid, A. P. 1982, Journal of the American Statistical Association, 77, 605

\bibitem[]{}

 Eastman, J., Gaudi, B., Agol, E. 2013, PASP,125,83	

\bibitem[]{}

 Hartkopf, W. I., McAlister, H. A. \& Franz, O. G.  1989, AJ,98,1014

\bibitem[]{}

 James, F. 2006, Statistical Methods in Experimental Physics. (Singapore:
                     World Scientific Publishing Co.)

\bibitem[]{}

 Lucy, L.B. 2014, A\&A, 563, 126 (L14)

\bibitem[]{} 

 Muller, M.E. 1959, Comm. Assoc. Comp. Mach. 2, 19

\bibitem[]{}
 Press W.H., Teukolsky S.A., Vetterling W.T., Flannery B.P. 1992, Numerical
          Recipes (2nd Ed.). (Cambridge: Cambridge Univ. Press) 

\bibitem[]{}

 Schaefer, G. H., Simon, M., Beck, T. L., Nelan, E. \& Prato, L.	
                                2006, AJ, 132, 2618	


\end{thebibliography}
\end{document}